\documentclass[12pt,titlepage]{article}
\usepackage{latexsym}
\usepackage{amsfonts}
\usepackage[latin1]{inputenc}
\newcommand{\hs}{\hspace}
\newcommand{\cem}{\hspace{1cm}}
\newcommand{\hem}{\hspace{5mm}}
\newcommand{\tem}{\hspace{3mm}}
\newcommand{\vs}[1]{\rule[- #1 mm]{0mm}{#1 mm}}
\newcommand{\bA}{{\bar{A}}}
\newcommand{\bD}{{\bar{D}}}
\newcommand{\bF}{{\bar{F}}}

\newcommand{\bU}{{\bar{U}}}
\newcommand{\bV}{{\bar{V}}}
\newcommand{\bW}{{\bar{W}}}

\newcommand{\ovX}{{\overline{X}}}
\newcommand{\ovY}{{\overline{Y}}}

\newcommand{\cd}{{\cal D}}

\newcommand{\cl}{{\cal L}}

\newcommand{\cdb}{{\bar{\cd}}}

\newcommand{\al}{{\alpha}}

\newcommand{\dt}{{\delta}}
\newcommand{\eps}{{\epsilon}}
\newcommand{\vep}{{\varepsilon}}

\newcommand{\si}{{\sigma}}
\renewcommand{\th}{\theta}
\newcommand{\vp}{\varphi}
\newcommand{\om}{{\omega}}

\newcommand{\da}{{\dot{\alpha}}}

\newcommand{\thb}{\bar{\theta}}
\newcommand{\etab}{\bar{\eta}}
\newcommand{\vpb}{{\bar{\vp}}}
\newcommand{\sib}{{\bar{\sigma}}}

\newcommand{\xib}{{\bar{\xi}}}
\newcommand{\psib}{{\bar{\psi}}}
\newcommand{\chib}{{\bar{\chi}}}
\newcommand{\phib}{{\bar{\phi}}}

\newfont{\twelvemsb}{msbm10 scaled\magstep1}
\newfont{\eightmsb}{msbm8}
\newfont{\sixmsb}{msbm6}
\newfam\msbfam
\textfont\msbfam=\twelvemsb \scriptfont\msbfam=\eightmsb
\scriptscriptfont\msbfam=\sixmsb \catcode`\@=11
\def\Bbb{\ifmmode\let\next\Bbb@\else
  \def\next{\errmessage{Use \string\Bbb\space only in math mode}}\fi\next}
\def\Bbb@#1{{\Bbb@@{#1}}}
\def\Bbb@@#1{\fam\msbfam#1}
\newfont{\twelvegoth}{eufm10 scaled\magstep1}
\newfont{\fourteengoth}{eufm10 scaled\magstep2}
\newfont{\sixteengoth}{eufm10 scaled\magstep3}
\newfont{\tengoth}{eufm10}
\newfont{\eightgoth}{eufm8}
\newfam\gothfam
\textfont\gothfam=\twelvegoth \scriptfont\gothfam=\eightgoth
\def\frak{\ifmmode\let\next\frak@\else
  \def\next{\errmessage{Use \string\frak\space only in math mode}}\fi\next}
\def\frak@#1{{\fam\gothfam{{#1}}}}
\catcode`@=12

\newcommand{\sect}[1]{\setcounter{equation}{0}\section{#1}}
\newcommand{\ba}{\begin{array}}
\newcommand{\ea}{\end{array}}
\newcommand{\nn}{\nonumber}

\def\lp{\left(}
\def\rp{\right)}
\def\l[{\left[}
\def\r]{\right]}
\newcommand{\f}[2]{{\displaystyle\frac{#1}{#2}}}
\def\2#1{\mbox{#1}}
\def\s2{\sqrt{2}}
\def\is2{\frac{1}{\s2}}

\newcommand{\ie}{{\em {i.e.\ }}}

\newcommand{\prt}{\partial}
\newcommand{\dC}{\partial C}

\newcommand{\sid}{{\si^m \prt_m\ }}
\newcommand{\sibd}{{\sib^m \prt_m\ }}

\newcommand{\ci}[1]{\raise5pt\hbox{$\scriptstyle#1$}} % correct index

\newcommand{\loco}{{\mathop{ \, \rule[-.06in]{.2mm}{3.8mm}\,}}} %proj. on lowest comp.

\begin{document}
%\fnsymbol{footnote}
\begin{titlepage}

\begin{center}
{\LARGE {\bf Correspondence between the 3-form and a
non-minimal  multiplet in supersymmetry
\\[3.mm]
}} \vskip 1cm

{\large G. Girardi} \footnote{on leave of absence of LAPTH, UMR
5108 du CNRS associ\'ee au L.A.P.P.},
{\large R. Grimm} \\

{\em CPT \footnote{UMR 6207 du CNRS}, C.N.R.S. Luminy, Case 907,
F-13288 Marseille Cedex 09, France} \\[1.2cm]

{\large B. Labonne and J. Orloff} \\
{\em LPC, Universit\'{e} Blaise Pascal, F-63177, Aubi\`{e}re Cedex, France}\\
%{\bf{Version du 17/11}}
\end{center}
\vs{15}

\centerline{ {\bf Abstract}}

\indent In analogy to the chiral-linear multiplet correspondence we establish a relationship between the 3-form (or gaugino condensate) multiplet and a coupled non-minimal $(0,1/2)$ multiplet, illustrated by a simple explicit example.

\vfill
%\rightline{LAPTH \ 755/99}
\rightline{CPT-P92-2007}
\rightline{PCCF-RI-0705}
\vskip .2cm \rightline{\today}
\end{titlepage}
%%%%%%%%%%%%%%%%%%%%%%%%%%%%%%%%%%%%%%%%%%%%%%%%%%%%%%%

\newpage
\pagenumbering{roman} %\tableofcontents
%\arabic{footnote}
%%%%%%%%%%%%%%%%%%%%%%%%%%%%%%%%%%%%%%%%%%%%%%%%%
%%%%%%%%%%%%%%%%%%%%%%%%%%%%%%%%%%%%%%%%%%%%%%%%%
%\newpage
\pagenumbering{arabic}
\section{Introduction}
%\subsec{ General remarks \label{3F1}
 \cem The scalar multiplet \cite{WZ74c}, commonly termed chiral
multiplet, is the most popular realization of the $(0,1/2)$
representation (massive or massless) of supersymmetry in terms of local quantum
fields. It contains as components a complex scalar, a Majorana
spinor and a complex scalar auxiliary field. Another realization of the
same representation (generally massless in this case) is provided
by the linear multiplet \cite{FZW74}, given in terms of
a real scalar, a 2-index antisymmetric tensor gauge field, a
Majorana spinor and no auxiliary field. Contrary to the previous
one the linear multiplet is a gauge multiplet.
 In classical Lagrangian field theory one can establish
\cite{Sie79} a certain correspondence between the chiral multiplet
and the linear multiplet, sometimes referred to as {\em{chiral-linear
multiplet duality}}, in particular in applications where the
linear multiplet incarnates a
{\em{dilaton-axion multiplet}}. In this note we would like to draw
attention to yet another couple of realizations of the $(0, 1/2)$
representation, the
{\em{3-form multiplet}} \cite{Gat81} and a {\em{non-minimal $(0,
1/2)$ multiplet}} \cite{GatesSieg}, including simple chiral
multiplet couplings.

The 3-form multiplet made of a 3-index antisymmetric tensor gauge field,
 a complex scalar, a Majorana spinor and a real auxiliary field may be
 understood as a {\em{further constrained chiral multiplet}}.
 It is the basic ingredient in the context of gaugino condensation,
  but is also relevant in the theory of supersymmetric gauge anomalies
  and in the description of curvature squared terms and Chern-Simons forms
 in supersymmetry. The non-minimal $(0, 1/2)$ multiplet, on the other
 hand, is less well known. In this note, we would like to outline
  a relation with the 3-form multiplet in very much the same vein as the above
   mentioned chiral-linear correspondence.
To be definite, we shall exhibit here a very simple toy model,
 coupling the gaugino-condensate multiplet to a single generic
 chiral multiplet and suggest a corresponding coupling  of the non-minimal
  $(0, 1/2)$ multiplet.

 It may be worthwhile to comment briefly on the  notion of {\em{gaugino condensate multiplet}}. In a supersymmetric gauge theory the gauge field-strength tensor is promoted to a multiplet containing as superpartners the gaugino and a real bosonic auxiliary field. The corresponding gaugino superfield, denoted $W_\al, \bW^\da$ is chiral ($ \bD^\da W_\al=0, D_\al \bW^\da=0$) and subject to the additional constraint $D^\al W_\al= \bD_\da \bW^\da$.
Irrespectively of the mechanisms underlying {\em{gaugino
condensation}}, the constraints on $W_\al,
\bW^\da$ imply that the {\em{condensate}} superfields $tr(W^\al W_\al) ,
tr(\bW_\da \bW^\da)$ are not only chiral
but fulfill the additional condition $$D^2\,tr(W^\al W_\al) -
\bD^2\,\, tr(\bW_\da \bW^\da)= i\vep_{klmn} tr \lp F^{kl}
F^{mn}\rp.$$ Interestingly enough, this supermultiplet can be
viewed as a particular realization of a generic 3-form gauge
theory in superspace with the 3-form gauge potential related to
the Chern-Simons form of the Yang-Mills theory.

\sect{ The 3-Form Multiplet  \label{3F2}}

\cem In multiplets of supersymmetry different components may be
assigned different $R$-weights, in relation to their supersymmetry
transformations and the chiral properties of their generators
\cite{FF77}. As it seems reasonable to assign vanishing $R$-weight
to gauge potential components, the $R$-weights of their
supersymmetry partners are then determined correspondingly.
Precisely in the case of $C_{klm}(x)$, the 3-form gauge potential
of the gaugino condensate multiplet with vanishing $R$-weight, the weights of
the other components are dictated by supersymmetry: in
units where $r(\theta)=r(\bD)=+1,\, r(\bar{\theta})=r(D)=-1$, the
complex scalar $Y(x), \ovY(x)$ has $r(Y)=+2,\, r(\ovY)=-2$. The fermionic components $\eta_\al(x), \etab^\da(x)$ acquire $r(\eta)=+1, r(\etab)=-1$, whereas
$H(x)$, the real auxiliary field has $r(H)=0$. Therefore, $H(x)$ may constitute by itself an R-inert supersymmetric Lagrangian in
analogy with the Fayet-Iliopoulos $D$-term familiar in
supersymmetric gauge theory.

In superfield language, the 3-form multiplet is characterized by
the superfields $Y, \ovY$ subject to the chirality conditions
\begin{equation}\label{ChiY} \bD^\da Y \,=\,0 , \hem D_\al \ovY \,=\,0 ,
\end{equation}
and the additional constraint
 \begin{equation}\label{conY} D^2 Y
- \bD^2 \ovY \,=\,  -8i\, \dC (x),
\end{equation}
with $\dC=-\frac{4}{3} \eps^{klmn}\prt_k C_{lmn}$. These superfield relations
have an interpretation as Bianchi identities in superspace geometry \cite{Gat81}, \cite{BPGG96}.
Component fields are identified as usual by projection to lowest superfield components
\begin{equation}
\ovY{\loco} \,=\,\ovY(x), \tem \bD^\da \ovY{\loco} \,=\,\s2 \,
\etab^\da (x), \tem Y{\loco} \,=\,Y (x), \tem D_\al Y{\loco}
\,=\,\s2 \, \eta_\al(x),
\end{equation}
\begin{equation}
D^2 Y{\loco} + \bD^2 \ovY{\loco} \,=\,-8 \, H(x), \cem C_{lmn} \loco =C_{lmn}(x).
\end{equation}

Supersymmetry transformations for these components read

\begin{equation}
 \delta_{\xi} C_{mlk} \,=\,\frac{\s2}{16} \left(
\xib \sib^n \eta - \xi \si^n \etab \right) \vep_{nmlk},
\end{equation}
\begin{equation}
 \dt_{\xi} Y \, = \, \s2 \, \xi^\al \eta_\al , \cem
    \dt_{\xi} \ovY \, = \, \s2 \, \xib_\da \etab^\da ,
\end{equation}
\begin{equation}
 \dt_{\xi} \eta_\al = \s2 \xi_\al \lp H +i \dC \rp
  +i\s2 (\xib\sib^m \eps)_\al \prt_m Y,
\end{equation}
\begin{equation}
   \dt_{\xi} \etab^\da = \s2 \xib^\da \lp H -i \dC \rp +i \s2
(\xi \si^m \eps)^\da \prt_m \ovY,
\end{equation}
\begin{equation}
 \dt_{\xi} H = \f{i}{\s2} (\xib\sib^m\prt_m \eta)
+\frac{i}{\s2} (\xi \si^m \prt_m \etab) . \label{delH}
\end{equation}

Taking care of the overall $R$-weights of $Y$ and $\ovY$,
invariant component field Lagrangians may be obtained from
("$D$-term integration")
\begin{equation}\label{LYY}
    \cl_{Y \ovY} = \int d^2 \th d^2 \thb \,(Y \ovY) =
    -\prt_m Y \prt^m \ovY +\frac{i}{2}(\prt_m \eta \sigma^m
    \etab - \eta \sigma^m \prt_m \etab) +H^2 +{\dC}^2,
\end{equation}
 the kinetic Lagrangian density and ("$F$-term integration")
\begin{equation}\label{Hterm}
\int d^2 \th \, Y +\int d^2 \thb\, \ovY= 2H(x),
\end{equation}
giving rise to the $H$-term referred to above.

Let us consider, as a very simple example, the coupling to a single
chiral superfield\footnote{Component fields for $ \phi$ are
   $\phi \loco =A(x), D_\al \phi \loco =\s2 \chi_\al (x), D^2
  \phi \loco =-4F(x)$, likewise for $\phib$. }, $\phi$, of vanishing $R$-weight, \ie adding a
kinetic density $$ \int d^2 \th d^2 \thb \, \phi \phib,$$ and
generalizing (\ref{Hterm}) to
$$ \int d^2 \th \, Y \, U
(\phi) +\int d^2 \thb \, \ovY \, \bar{U} (\phib), $$ with $U(\phi),
\bar{U}(\phib)$ at most quadratic in the renormalizable case.

In fact, using the explicit solutions of
the constraints (\ref{ChiY}, \ref{conY})
\begin{equation}\label{omega}
    Y= -4  \bD ^2\ \Omega, \tem  \ovY= -4  D ^2\ \Omega,
\end{equation}
with $\Omega$ the real unconstrained pre-potential (undetermined up
to a linear superfield pre-gauge transformation) and employing
integration by parts in superspace, the complete  action density
may be written as a pure $D$-term integration

\begin{equation}\label{complact}
 \int d^2 \th d^2 \thb \ \left[ \frac{}{} Y \ovY + \phi \phib + 16 \Omega \lp U(\phi)
 +\bU (\phib) \rp \right],
\end{equation}
with suitable superfield equations of motion.

\sect{A simple model }

\cem In this section, we consider a particular combination of the
3-form and a chiral multiplet in  choosing  $U(\phi)=\al +\mu
\phi, \, \mu \in \mathbb{R}$, giving rise to the superfield action
density:
\begin{equation}\label{Laggen}
 \int d^2 \th d^2 \thb  \lp  \phi \phib +Y \ovY \rp
 +\int d^2 \th  \lp \al +\mu \phi \rp Y
  + \int  d^2 \thb \lp {\bar{\al}}+\mu \phib \rp \ovY.
\end{equation}
At the component field level, this action contains the kinetic
terms for  $A, \bA, \chi, \chib$ (chiral multiplet) and $Y, \ovY,
\eta, \etab$ (3-form multiplet), mixing terms of these with $F,
\bF$ (chiral multiplet), $H$ (3-form multiplet) and, last but not
least, the terms containing $\dC$, the field-strength of the
3-index antisymmetric gauge potential.

In many cases, in supersymmetric field theories, {\em{elimination
of auxiliary fields}} means rather  diagonalization in terms
 of non propagating fields (no derivative terms in the action
 density) with trivial algebraic equations of motion. In the case
 at hand, this can be done easily for the part of the action
 density containing $F, \bF$ and $H$, yielding
\begin{eqnarray}\label{LagY}
\cl&=&-\prt^m \! A \, \prt_m \bA - \f{i}{2} \lp \chi \si^m \prt_m
\chib + \chib \sib^m \prt_m \chi \rp -\prt_m Y \prt^m \ovY
-\frac{i}{2}( \eta\sigma^m \prt_m \etab + \etab\sib^m  \prt_m
\eta) \nn \\&&
 - \mu\,\lp \chi \eta + \chib  \etab  \rp-|\,\mu\,Y|{}^2
 -|\al+\mu \,A|{}^2+\left[ \dC +\frac{i}{2}\lp \al -\bar{\al}+\mu\,(A -\bA)\rp \right]^2 \nn
 \\&& +\, \cal{F}\, \overline{\cal{F}}+ {\cal{H}}{\cal{H}},
\end{eqnarray}
with diagonalized auxiliary fields
 \begin{equation} \label{auxeq}{\cal{F}}=F+\mu \ovY, \tem \overline{\cal{F}}=\bF+\mu
Y,\tem{\cal{H}}=H+\f12 (\al+\bar{\al})+ \frac{\mu}{2}(A + \bA).
\end{equation}

The complex scalar $Y, \ovY$ satisfies a Klein-Gordon equation
with mass $\mu$, the Weyl spinors $ \eta,\chib,$
combine into a Dirac spinor of the same mass. The equations of
motion for the fields $A, \bA, C_{klm}$ are most
conveniently written using
 $A =A_1 +i A_2,\, \al=\al_1 +i \al_2$, so that
\begin{eqnarray} \label{aplus}
 \Box A_1 - \mu^2\,\lp A_1 +\frac{\al_1}{ \mu} \rp&=& 0,  \\ \label{amoins}
 \Box A_2 -  \mu \dC  \,&=&0, \\
 \label{dc}
\prt_m \, ( \dC - \mu \, A_2 ) &=&0.
 \end{eqnarray}
The last  equation is compatible with a constant $ K_2 =\dC - \mu \, A_2 $,
giving rise to a shifted Klein-Gordon equation for $A_2$
\begin{eqnarray} \label{KGA}
&& (\Box- \mu^2 )\,\lp A_2 +\frac{ \al_2+K_2}{\mu} \rp  =0.
 \end{eqnarray}
We would like to stress that these features arise necessarily in
the context
   of models dealing with gaugino condensation.

%\newpage
 \sect{ The $X-Y$ Correspondence}

\cem Independently of supersymmetry, the 3-index antisymmetric
gauge potential $C_{klm}$ has been employed in the context of the
cosmological constant problem \cite{Haw84}, \cite{Duf89}. The
derivative quadratic action density is proportional to $ (\dC)^2$.
This density can be related to a constant considering the density
$$ X^2 +X \dC$$ with $X(x)$ a real field. Varying with respect to
$X$ and substituting back reproduces $(\dC)^2$. On the other hand,
varying with respect to $C_{klm}$ implies $\prt_m X=0$, \ie $X$ a
constant.

 This mechanism can be extended to the supersymmetric
case, \textit{e.g.} the 3-form multiplet. Here we consider the combination
\begin{equation}
\int d^2\theta  d^2\thb \left[ \frac{}{} -X \ovX  -X Y -\ovY \ovX
+ 16\Omega \lp U(\phi)+\bU(\phib) \rp +\phi \phib \  \right].
\end{equation}
$X, \ovX$ is a complex unconstrained superfield, $Y,\ovY$, the
3-form superfield introduced above and $\Omega$ its unconstrained
real pre-potential. $\phi, \phib$ are considered as  spectator
superfields. Varying with respect to $X, \ovX$ just implies $
X=\ovY,\, \ovX=Y$ and one recovers (\ref{complact}) upon
substitution. As to variation with respect to the 3-form multiplet
we shall use the solution (\ref{omega}) of the constraints and
integration by parts in superspace to arrive at
\begin{equation}
\int d^2\theta  d^2\thb \left[ \frac{}{}- X \ovX +4 \Omega \lp \bD^2 X +D^2
\ovX + 4  U(\phi)+ 4\bU(\phib) \rp +\phi \phib \ \right],
\end{equation}
where $\Omega$ may be considered as a Lagrange multiplier superfield giving rise to a constraint
\begin{equation}\label{Xcon}
 \bD^2 X +D^2 \ovX + 4U(\phi)+4\bU(\phib)=0,
\end{equation}
 that can be separated into  two constraints
\begin{equation}\label{conX}
  {\bD}^2 X = -4U(\phi)-4 K , \cem {D}^2 \ovX=-4 \bU (\phib)-4 \bar{K},
\end{equation}
 related by a constant $K=- \bar{K} =iK_2$, which might be
absorbed in a redefinition of $U(\phi), \bU(\phib)$.
 In other words, in supersymmetry, the analogue of the constant
 mentioned above (in the non supersymmetric case) is given by a complex
  superfield, $X,\ovX$.

The component
field action is then obtained from
\begin{equation}\label{lXphi}
\int d^2\theta  d^2\thb \left[ \frac{}{} - X \ovX +\phi \phib \ \right].
\end{equation}
  In the
case $U=0$, this multiplet has been presented in \cite{GatesSieg}.
We shall call it {\em{non-minimal}} in what follows and use the
term {\em{coupled non-minimal}} in the case of non vanishing $U$,
to be discussed in the next section.
%\newpage
\section{The coupled non-minimal multiplet}

\cem The superfield constraints (\ref{conX}) determine a multiplet
of 12 bosonic and 12 fermionic component field degrees of freedom,
identified as usual by successive applications of covariant spinor
derivatives. We define the  component fields contained in $X$ as
\begin{eqnarray}\label{locoX}
 X \loco &=& X, \cem  D_\al X\loco = \s2 \,{\psi}_\al,  \hem \ \
{\bD}^\da X\loco = -\s2 \, {\bar{\omega}}^\da,\hem
  \nn  \\
 {\bD}^\da  D_\al   X \loco &=& {V_\al}^\da,\hem {\bD}^\da D^2 X \loco
  =-4\bar{\rho}^\da,\hem  D^2 X \loco = -4E.
\end{eqnarray}
Observe that the $\thb^2$ component is given in terms of $A$, Cf.(\ref{conX}). For
$\ovX$ we define similarly
\begin{eqnarray}\label{locoXb}
 \ovX \loco &=& \ovX, \cem  D_\al \ovX\loco = -\s2 \,{\omega}_\al,  \hem \ \
{\bD}^\da \ovX\loco = \s2 \, {\bar{\psi}}^\da,\hem
  \nn  \\
 D_\al {\bD}^\da    \ovX \loco &=& {{\bV}_\al}{}^\da,\hem {D}_\al \bD^2 \ovX \loco
  =-4{\rho}_\al,\hem  \bD^2 \ovX \loco = -4 \bar{E}.
\end{eqnarray}
Projecting $\frac{1}{16}D \bD^2 D ( -X \ovX)$ to lowest
components gives the canonical component field action density\footnote{Primes indicate derivatives
with respect to $A$ or $\bA$, as the case may be.}
\begin{eqnarray}\label{XX}
\cl_X
&=&- \prt_m X \prt^m \ovX -\f{i}{2} \lp \om \sid \bar{\om} + \bar{\om} \sibd\om \rp  \\
 &&- |\,U|^2
- U' \lp \ovX F +\om \chi \rp-  {\bU}' \lp X
\bF +\bar{\om} \chib \rp +\frac{1}{2}
\,\ovX U'' (\chi \chi) +\frac{1}{2}\,X {\bU}''(\chib \chib) \nn \\
&&- \frac{1}{2} V_m \bV^m - E \bar{E} - \f{i}{2} \psi\lp \sid
\psib +i \s2 \rho \rp
- \f{i}{2}\psib \lp\sibd\psi+i \s2 \bar{\rho}\ \rp ,\nn
\end{eqnarray}
describing a complex scalar $X$ and a Majorana spinor $\om$ as
physical fields. Auxiliary fields consist of a complex scalar $E$,
a complex vector $V_m$, and 2 Majorana spinors $\psi, \rho$. This
action density is invariant under  supersymmetry transformations:
\begin{eqnarray}\label{suztf}
&&\delta_{\xi} X = \s2 \ (\xi \psi - \xib
{\bar{\omega}}),\hs{0.8cm}
 \delta_{\xi} \ovX  = \s2\ ( \xib {\bar{\psi}}-\xi \omega),\nn \\
&&\delta_{\xi} {\psi}_\al= \s2 E\ \xi_\al  -\frac{1}{ \s2}V_m \
(\xib \sib^m \eps)_{\al} ,  \hs{0.8cm} \delta_{\xi} {\psib}^\da =
\s2 \bar{E}\ \xib^\da  -\frac{1}{ \s2}\ \bV_m (\xi
\si^m \eps)^\da, \nn \\
&&\delta_{\xi} {\bar{\omega}}^\da = -\s2 U  \xib^\da -\frac{1}{
\s2} (V_m +2i \prt_m X)(\xi \si^m \eps)^\da ,\nn \\ &&\delta_{\xi}
{\omega}_\al = -\s2 \bU  \xi_\al
-\frac{1}{ \s2}\ (\bV_m +2i \prt_m \ovX)(\xib \sib^m \eps)_{\al},\nn \\
&&\delta_{\xi} \ V_m = (\xi \si_{m}\bar{\rho})+\s2 i \ (\xi
\si_{n} \sib_m\prt^n \psi -\xib \sib_{m}\si_n \prt^n \bar{\omega})
-(\xib \sib_m \chi)U', \nn \\
&&\delta_{\xi} \ \bV_m = (\xib \sib_{m}{\rho})+\s2 i \ (\xib \sib_{n}\si_m \prt^n \bar{\psi}-\xi
\si_{m}\sib_n \prt^n \omega)
-(\xi \si_m \chib)\bU',\nn \\
&&\delta_{\xi} \bar{\rho}^\da = 2i\  \ \prt_m E \ (\xi \si^m
\eps)^\da + (2i\prt_m V^m
-2 \Box X -U'' \chi \chi +2 U' F)\ \xib^\da , \nn \\
 &&\delta_{\xi} {\rho}_\al = 2i\ \prt_m \bar{E}\ (\xib \sib^m
\eps)_\al \
 + (2i\prt_m \bV^m -2 \Box \ovX -\bU'' \chib \chib +2 \bU' \bF)\
 \xi_\al\ ,\nn \\
&&\delta_{\xi} E = \xib\, \bar{\rho}, \hs{0.8cm}\delta_{\xi}
\bar{E} = \xi\, {\rho},\nn \\ &&\delta_{\xi} \bD^2 X =-4 \s2 U'
(\xi \chi ),\hs{0.8cm} \delta_{\xi} D^2 \ovX = -4 \s2 \bU' (\xib
\chib ).
\end{eqnarray}

Adding the kinetic Lagrangian for $\phi$
\begin{eqnarray}\label{lska}
{\cal{L}_S} &=&-\prt^m \! A \, \prt_m \bA - \f{i}{2} \lp \chi
\sid \chib + \chib \sibd\chi \rp +F \bF,
\end{eqnarray}
the complete Lagrangian is
\begin{eqnarray}\label{LL}
\cl &=&-\prt^m \! A \, \prt_m \bA - \frac{i}{2}\lp \, \chi \sid \chib
+\chib \sibd \chi \rp- \prt_m X \prt^m \ovX  -\frac{i}{2}\lp \, \om \sid
\bar{\om}+ \bar{\om} \sibd \om \rp
\ \nn \\
&&-  U'(A)\,  \om \chi - {\bU}'(\bA) \, \bar{\om} \chib - |U'(A)|^2 X \ovX  -|U(A)|^2 \nn \\&&  +\frac{1}{2}
\,\ovX U''(A)(\chi \chi )+\frac{1}{2}\,X {\bU}''(\bA) (\chib
\chib) + {\cal{F}}\,
{\overline{\cal{F}}}-\frac{1}{2} V_m \bV^m - E \bar{E}\nn \\ &&-
\f{i}{2} \psi\lp \sid \psib +i \s2 \rho \rp - \f{i}{2}\psib
\lp\sibd\psi+i \s2 \bar{\rho}\ \rp,
\end{eqnarray}
with
 \begin{equation} \label{auxeqX}{\cal{F}}=F- X \bU ', \tem \overline{\cal{F}}=\bF- \ovX U', \end{equation}
and exhibiting the general scalar potential
\begin{equation}
{\cal{V}} =|U'(A)|^2 X \ovX  +|U(A)|^2.
\end{equation}
In order to make contact with the simple model of section 3, we
set $U(A)= \al +\mu A,\, \bU(\bA)= \bar{\al}+\mu \bA$. Then (\ref{LL}) describes two  complex scalar fields and a Dirac field with common mass $\mu$, just like the Lagrangian
(\ref{LagY}).
The difference between the two Lagrangians  appears in the auxiliary  field sector and, correspondingly, in the component field  supersymmetry transformations. Moreover, it should be stressed that $Y, \ovY$ represents a gauge multiplet, whereas $X, \ovX$ does not; this is also the case for the linear-chiral multiplet correspondence.

\section{Conclusions}

\cem The main purpose of this short communication was to establish
a correspondence between the 3-form multiplet and a non-minimal
multiplet, in analogy to the well-known relation between the
2-form (\ie  linear) multiplet and the chiral multiplet. Observe
that in both cases the correspondence can only be established
under certain restrictive assumptions.

Although the 3-form multiplet and the non-minimal multiplet might be considered as {\em{exotic multiplets}}, they are not. As indicated in the introduction, the 3-form multiplet  describes naturally the gaugino squared chiral superfield $ tr \lp W^\al W_\al \rp$ and its complex conjugate.
On the other hand, the non-minimal multiplet appears naturally in the context of the solution of the chiral superfield constraints, \ie $\phi\, =\, \bD^2 \varphi, \, \phib \, =\, D^2 \vpb $, in terms of unconstrained potentials $\vp, \vpb$, defined up to pre-gauge transformations $\vp \rightarrow \vp+\xi, \, \vpb \rightarrow \vpb+\xib$. These superfields are themselves subject to the pre-constraints $\bD^2 \xi =0,\, D^2 \xib=0$, leaving $\phi,\phib$ invariant.

Let us mention as well that the above-mentioned 3-form constraints appear in an intriguing way in supergravity, in the framework of $U(1)$ superspace. The chiral supergravity superfields $R, R^\dagger$ are intertwined with the vector superfield $G_a$ through the relation $\cd^2 R -\cdb^2 R^\dagger =4i \cd^a G_a$.
Remarkably enough, here, the {\em{$H$-term }} of $R, R^\dagger$ corresponds to a {\em{$D$-term}} of the $U(1)$ supergravity sector.

The emphasis of the present note was to draw attention to the basic features of the correspondence between the 3-form multiplet and the non-minimal multiplet restricting ourselves to quite elementary considerations. More involved structures as well as the corresponding supergravity couplings will be the subject of forthcoming publications.

\addcontentsline{toc}{section}{\bf  REFERENCES}
\bibliography{gglobib}
\bibliographystyle{unsrt}

\end{document}